\documentclass[manuscript=article,layout=twocolumn]{achemso}
\setkeys{acs}{keywords = true}
\setkeys{acs}{usetitle = true}
\usepackage[version=3]{mhchem} 
\usepackage{color}
\usepackage{threeparttable}

\usepackage{lineno}
\usepackage{hyperref}
\usepackage{multicol}
\usepackage{blindtext}

\usepackage{xpatch}
\usepackage{blindtext}

\makeatletter
\xpatchcmd{\titlepage}{\@restonecolfalse\newpage}{\@restonecolfalse}{}{}
\xpatchcmd{\endtitlepage}{\if@restonecol\twocolumn \else \newpage \fi}{\if@restonecol\twocolumn \else  \fi}{\typeout{success}}{\typeout{fail}}
\makeatother
\onecolumn

\author{Zhen Zeng}
\affiliation{State Key Laboratory of Engines, Tianjin University, Tianjin, 300350, China.}
\author{Kai Sun}
\affiliation{State Key Laboratory of Engines, Tianjin University, Tianjin, 300350, China.}
\author{Rui Chen}
\affiliation{State Key Laboratory of Engines, Tianjin University, Tianjin, 300350, China.}
\author{Mengshan Suo}
\affiliation{State Key Laboratory of Engines, Tianjin University, Tianjin, 300350, China.}
\author{Zhizhao Che}
\email{chezhizhao@tju.edu.cn}
\affiliation{State Key Laboratory of Engines, Tianjin University, Tianjin, 300350, China.}
\alsoaffiliation{National Industry-Education Platform of Energy Storage, Tianjin University, Tianjin, 300350, China.}
\author{Tianyou Wang}
\email{wangtianyou@tju.edu.cn}
\affiliation{State Key Laboratory of Engines, Tianjin University, Tianjin, 300350, China.}
\alsoaffiliation{National Industry-Education Platform of Energy Storage, Tianjin University, Tianjin, 300350, China.}
\title{Correlation between 2D Square Ice and 3D Bulk Ice by Critical Crystallization Pressure}


\begin{document}

\begin{abstract}
Low-dimensional ice trapped in nanocapillaries is a fascinating phenomenon and is ubiquitous in our daily lives. As a decisive factor of the confinement effect, the size of nanocapillary significantly affects the critical crystallization pressure and crystalline structure, especially for multi-layered ices. By choosing square ice as a typical two-dimensional (2D) multi-layered ice pattern and using all-atom molecular dynamics simulations, we further unveil the variation mechanism of critical crystallization pressure with the nanocapillary size. The results show a strong dependence of the critical crystallization pressure on the size of the graphene sheet for monolayer, bilayer, and trilayer square ice. The quasi-macroscopic crystallization pressure, the actual pressure of water molecules, and the freezable region between them are all strongly dependent on the nanocapillary width. As the size of the capillary becomes larger in all three directions, the critical crystallization pressure converges to the true macroscopic crystallization pressure, which is very close to the value of the crystallization pressure for bulk ice. A direct correlation is established between 2D square ice and three-dimensional (3D) bulk ice by the critical crystallization pressure. There is an unfreezable threshold for crystallizing spontaneously in practice when the quasi-macroscopic crystallization pressure is equal to the actual pressure, which can explain the limit of nanocapillary width for multi-layered ice.
\end{abstract}


\section{Introduction}\label{sec:sec1}
The freezing of water is one of the most common and vital physical processes on the planet, and also ubiquitous in our daily life\cite{Mishima1998SupercooledGlassyWater}. At least 19 bulk ice phases have been confirmed by experiments so far, and more phases have been revealed from simulation predictions\cite{Debenedetti2003SupercooledGlassyWater, gasser21, Huang2016UnderNegativePressure, Poole1992MetastableWater, Sciortino1995CrystalStabilityLimits, Zhao2014TwoDimensionalAmorphousIce, Zheng1991HomogeneousNucleationLimit}. Different from the bulk ice that is most familiar to us, water molecules trapped in nano spaces can form novel crystalline structures due to the confinement effect. The confined crystal and quasicrystal structures of water molecules have been receiving increasing attention over the past two decades, yielding many different characteristics and practical applications\cite{Compton2012MechanicalPropertiesofGraphene, he19, kwac17, Neek2016ViscosityofNanoconfinedWater, wang19}.

There have been plenty of low-dimensional ice phases revealed by classical and ab-initio molecular simulations, including the quasi-one-dimensional crystal structures and two-dimensional (2D) multi-layered ices. For instance, single-walled ice nanotubes with square\cite{Koga2000IceNanotube}, pentagonal\cite{Bai2003PentagonandHexagonIce, koga01, Koga2000IceNanotube}, hexagonal\cite{Bai2003PentagonandHexagonIce, koga01, Koga2000IceNanotube}, and heptagonal\cite{Bai2003PentagonandHexagonIce, Byl2006UnusualHydrogenBonding} structures, and multi-walled ice nanotube with helical structure\cite{Bai2006MultiwalledIceHelixes} under higher hydrostatic pressure are typical quasi-one-dimensional crystal structures. Monolayer ices with octagonal\cite{Bai2010MonolayerClathrate}, tetragonal\cite{Bai2010MonolayerClathrate}, hexagonal\cite{Zhao2014FerroelectricHexagonal}, flat rhombic\cite{Zhao2014FerroelectricHexagonal}, puckered rhombic\cite{Bai2010MonolayerClathrate}, and square structures\cite{algara-siller15}, and multilayer ices with the same structures\cite{Chen2017DoubleLayerIce, Corsetti2016HighDensityBilayer, Koga2005HydrophobicSurfaces, Koga2000AmorphousPhases} and some other structures\cite{Bai2012PolymorphismandPolyamorphism, Zhu2015CompressionLimit} are representative 2D multi-layered ices. There are also some new crystalline structures that have been confirmed experimentally, such as polygonal nanotube ice structures by systematic X-ray diffraction analysis\cite{Maniwa2005PentagonaltoOctagonal} and 2D square ice at room temperature by transmission electron microscopy\cite{algara-siller15}. The simulation and experiment studies on low-dimensional ice also provide widespread applications in many fields\cite{garcia22, lin19, negi22}.

Compared to macroscopic bulk ice, the major difference for the low-dimensional ice is the effect of confinement, which can be dominated by capillarity in large pores and adsorption in minuscule pores\cite{Zhang2018UnfreezableThreshold}. The confinement effect works directly on the confined water molecules and causes ultrahigh pressure, resulting in plenty of novel crystalline structures\cite{algara-siller15}. Therefore, the high van der Waals pressure is the most important factor for the formation of low-dimensional ice\cite{algara-siller15}, as crucial as temperature to the freezing process for bulk ice. In recent years, many studies about low-dimensional ice focused on the effects of the type of confining surfaces\cite{bampoulis16, goswami20, kim16, Qiu2015InhomogeneousNanoconfinement, Ruiz2018FlexibleConfiningSurfaces}, the thermodynamic variables\cite{Zhu2016AbStacked, Zhu2016TrilayerIces, Zhu2016BucklingFailureofSquareIce, Zhu2017SuperheatingofMonolayerIce}, and the size of nanocapillary\cite{Zhu2015CompressionLimit, Zhu2016TrilayerIces}. It has been proven that the size of nanocapillary significantly affects the crystallization pressure and crystalline structure, especially for multi-layered ices. However, current studies on the size of nanocapillary are mainly about the separation of confinement surface, i.e., the width of the nanocapillary. Typically, Zhu et al.\cite{Zhu2015CompressionLimit} performed a detailed simulation study and summarized a compression-limit phase diagram of 2D water, by characterizing the compression limit from a mechanics point of view. However, the stability limit of the multilayer ice results in a very small range for the change of nanocapillary width (from 6.0 to 11.6 {\AA}), and determines the discontinuity of the diagram. Due to the importance of crystallization pressure for the low-dimensional ice, the influence of nanocapillary size in all three directions needs to be studied further.

Among the plenty of novel crystalline structures formed in nanoscale confinements, 2D square ice (including nearly square ice) is noteworthy because the square pattern is qualitatively different from the conventional tetrahedral structure and has been confirmed experimentally\cite{algara-siller15}. In this study, we choose square ice as a typical 2D multi-layered ice pattern and further explore the effect of nanocapillary size on ice formation. By all-atom molecular dynamics simulations, we change the size of nanocapillary in all three directions and unveil the variation mechanism of critical crystallization pressure with the nanocapillary size. This study not only explains why the two-dimensional ice has a strict limit on the number of layers\cite{Zangi2004WaterConfinedReview}, but more importantly, establishes a direct correlation between 2D square ice and 3D bulk ice by critical crystallization pressure.

\section{Computational methods}\label{sec:sec2}
\subsection{Configuration of simulation system}
\begin{figure}[t]
  \centering
  \includegraphics[width=0.6\columnwidth]{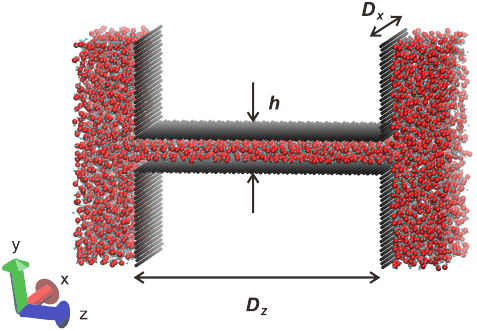}\\
  \caption{Configuration of the MD simulation system. The red, cyan, and black beads indicate oxygen, hydrogen, and carbon atoms, respectively.}\label{fig:fig1}
\end{figure}

Figure \ref{fig:fig1} is the initial configuration of the MD simulation system, which is similar to our previous studies \cite{Zeng2022PartitionedSquareIce}. A capillary is formed by two parallel graphene sheets and on either side of it are two water reservoirs with each containing the same amount of water molecules. To facilitate the comparison of results, especially in the study of structural stability, we fixed the aspect ratio of the computing domain for graphene sheets. The size of the graphene sheet ($D_x$ in $x$ direction and $D_z$ in $z$ direction as shown in Figure \ref{fig:fig1}) is changed proportionally and chosen as 26.6 {\AA} $\times$ 21.9 {\AA}, 42.5 {\AA} $\times$ 35.0 {\AA}, 68.0 {\AA} $\times$ 56.0 {\AA}, 108.8 {\AA} $\times$ 89.6 {\AA}, respectively. The width of the nanocapillary ($h$ in the $y$ direction) has a remarkable influence on the structure of the confined water molecules. Our previous results show that 6.5 {\AA} is the appropriate width for the formation of the flat monolayer square ice\cite{Zeng2022PartitionedSquareIce}. As the width increases, the flat monolayer square ice is puckered and squeezed into two layers. Similarly, 9.0 and 11.5 {\AA} are the appropriate widths for the formation of bilayer and trilayer square ice. Hence, we chose the typical widths (i.e., 6.5 {\AA}, 9.0 {\AA}, 11.5 {\AA}, and 14.0 {\AA}) to accommodate one, two, three, and four layers of water molecules, respectively. The total number of water molecules in our study is set to be 960, 1680, 2640, 4320, and 6960, respectively, for different sizes of graphene sheets.

\subsection{Molecular dynamics simulation}
 All MD simulations are performed in the isothermal, isobaric ensemble, in which the pressure ($P_z$) and the temperature ($T$) are controlled by the Nos\'{e}-Hoover barostat and thermostat. There is a constant temperature ($T$ = 298 K) during the simulations. To examine various lateral pressure-dependent states, the range of the lateral pressure $P_z$ is varied from 0 to 4.0 GPa. We used the anisotropic barostat in the pressurization simulations, which means the $x$, $y$, and $z$ dimensions are controlled independently. The $z$-length of the simulation box will change only according to the lateral pressure in the $z$ direction, and the box dimensions in the $x$ and $y$ directions do not change in our pressurization simulations. It is difficult to reproduce the terraced confinement in the MD analysis. Therefore, the high pressure induced by encapsulating graphene sheets is modeled to a first approximation by applying a hydrostatic pressure $P$ (which named lateral pressure) in the direction parallel to the graphene layers (in $z$ direction), and the lateral pressure $P$ is increased with different pressurization rates (0.12, 0.10, 0.08, and 0.02 GPa every 1 ns) for comparison in the pressurization process. Square ice can form independently for the rigid or flexible graphene confinement\cite{algara-siller15, Ruiz2018FlexibleConfiningSurfaces}, therefore, the carbon atoms in the graphene sheets are fixed in this study. Three directions in this study are all set to periodic boundary conditions. The time step is set to 1.0 fs for the velocity-Verlet integrator. In addition, the simulations last longer than 45 ns. The extended simple point charge (SPC/E) water model containing the short-range Lennard-Jones potential and the long-range Coulomb potential\cite{Berendsen1987EffectivePairPotentials} is used in this study. In the simulation of a multilayer square ice structure, the SPC/E water model is accurate by comparing with previous experimental results\cite{algara-siller15}. Other water model choices have also been tested for comparison in our previous studies, and the simulation shows the same results for the relation between graphene size and critical crystallization pressure, as shown in Figure S1 in the Supporting Information\cite{Zeng2022PartitionedSquareIce}. The 12-6 Lennard-Jones potential between the oxygen and the carbon atoms is used to model the water-carbon interaction. The energy and the distance parameters are $\epsilon$ = 0.114 kcal/mol and $\sigma$ = 0.328 nm\cite{Che2017SurfaceNanobubbles, Gordillo2000HydrogenBondStructure, Joly2011GiantLiquidSolidSlip, werder03}. In our simulations, the cutoff distance for the L-J potential is set to be 10.0 {\AA}. The particle--particle particle--mesh (PPPM) algorithm with an accuracy of $10^{-4}$ is used to compute the long-range interactions. The simulations are carried out by using LAMMPS\cite{Plimpton1995FastParallelAlgorithms}, and all the snapshots of simulation results are rendered in VMD\cite{Humphrey1996VMD}. Square icing parameters, $MCV_1$ and $MCV_2$, are used to quantitatively distinguish between the square ice structure and liquid water. The parameters can analyze the uniformity of the molecule's distribution in the direction of the hydrogen?oxygen bonds and spatial position, and identify the different water phases when we take appropriate thresholds.

\section{Results and discussion}
\subsection{Unfreezable threshold and quasi-macroscopic crystallization pressure for layered square ice}
\begin{figure}[t]
  \centering
  \includegraphics[width=0.9\columnwidth]{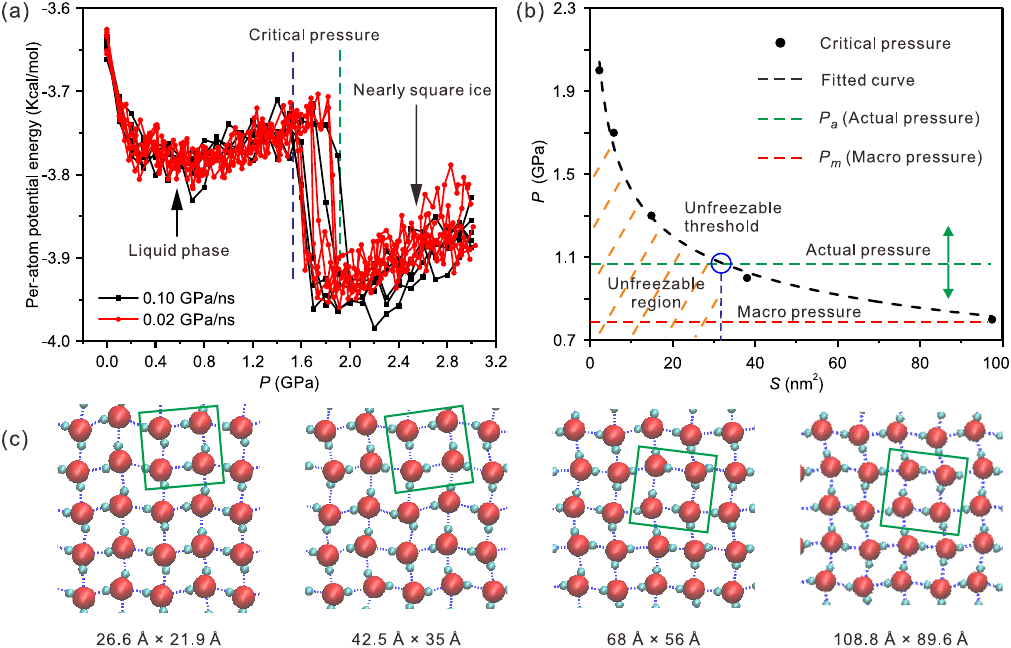}\\
  \caption{Results for the bilayer square ice ($h$ = 9.0 {\AA}). (a) Phase transformation of water molecules from liquid to nearly square ice with different pressurization rates (the size of graphene sheets is 26.6 {\AA} $\times$ 21.9 {\AA}). (b) Critical crystallization pressure as a function of graphene size. The black dots represent the characteristic critical crystallization pressure. The black dashed curve is the fitted curve of the characteristic critical crystallization pressure, and the horizontal green and red dashed lines represent the actual pressure and the quasi-macroscopic pressure, respectively. The vertical blue dashed line represents the unfreezable threshold for the graphene size, and the pink shaded area is the unfreezable region. (c) Top views of the simulation snapshots for graphene sheets of different sizes. The red and cyan globules indicate oxygen and hydrogen atoms, respectively, and the blue dashed lines represent hydrogen bonds. The green boxes mark square ice units to guide the eye.}\label{fig:fig2}
\end{figure}

 We carry out a series of all-atom MD simulations of different capillary widths and graphene sizes, and focus on the transition of water molecules from liquid to crystal in the pressurization process, especially the critical value of pressure for freezing. The critical crystallization pressure, under which the phase transition from liquid water to 2D ice occurs, is one of the most important criteria to determine whether the phenomenon of square ice happens. In the study of bilayer square ice ($h$ = 9.0 {\AA}), the lateral pressure $P$ is increased stepwise at a fixed rate of 0.10 GPa every 1 ns in the pressurization process, and the simulations for each graphene size are repeated six times by changing the initial velocity distributions of the water molecules while keeping the other settings identical. In this study, all the results of critical crystallization pressure are obtained in three approaches: (1) The transition is accompanied by a sudden drop of the potential energy, so the potential energy curve can be used as an intuitive estimation of identifying phase transitions. (2) The square icing parameters, $MCV_1$ and $MCV_2$, are calculated to quantitatively distinguish between square ice structures and liquid water\cite{Zeng2022PartitionedSquareIce, Zeng2023CriticalCrystallizationPressure}, and the same results of the critical crystallization pressure as the first method are obtained. (3) All the results of critical crystallization pressure are verified by observing the ice formation from simulation snapshots frame by frame. The results show that the critical crystallization pressure is not a fixed value for the same graphene size, but fluctuates within a certain range, as shown in Figure \ref{fig:fig2}(a). Moreover, there is a strong dependence of the fluctuation range of the critical crystallization pressure on the size of the confining graphene, and the fluctuation range narrows as the graphene sheets widen, as shown in Figure S2 in the Supporting Information. This phenomenon is similar to the range of temperature for the coexistence of nanoconfined ice and liquid\cite{Kastelowitz2018IceLiquidOscillations}.

By averaging the different values of the critical crystallization pressure, we can obtain a characteristic critical crystallization pressure for each graphene size. Figure \ref{fig:fig2}(b) summarizes the characteristic critical crystallization pressure as a function of graphene size, showing a strong dependence. We also summarize the critical crystallization pressure as a function of $D_z$, and the results show that there is no difference for the variation tendency between graphene size and critical crystallization pressure, as shown in Figure S3 in the Supporting Information for details. As the graphene sheet becomes larger, the critical crystallization pressure slowly decreases, converging to an approximately macroscopic crystallization pressure for bilayer 2D ice ($P_m$ in Figure \ref{fig:fig2}(b)). On the other hand, the actual pressure is the internal hydrostatic pressure of water molecules from the energy gain due to such squeezing effect, and the actual pressure of water molecules for $h = 9.0$ {\AA} ($P_a$ in Figure \ref{fig:fig2}(b)) can be estimated from the energy gain due to the squeezing effect ($P_a= E_a/d \approx 1.0$ GPa, where $E_a \approx 30$ meV/{\AA}$^2$ is the difference in the adhesion energy per unit area between graphene-water and graphene-graphene, and $d \approx 5.6$ {\AA} is calculated by subtracting the distance between the graphene plates without water molecules from the distance between the graphene plates filled with water molecules, i.e., the increased separation distance between the two graphene sheets because of the trapped water molecules\cite{algara-siller15}). Even though this method is just a rough estimate and the value of $P_a$ is not precise, this result shows that there is an unfreezable threshold for the graphene size when $P_a$ is equal to the characteristic critical crystallization pressure, and the curve of critical crystallization pressure between $P_a$ and $P_m$ represents the freezable region. The unfreezable threshold means it is hard to crystallize spontaneously in practice when the graphene sheet is smaller than the threshold, because the actual pressure is not high enough to match the critical crystallization pressure of icing. Considering that the results seem similar to the finite size effect convergence, we test a much slower pressure increasing rate (0.02 GPa every 1 ns) for comparison to dispel the concern about the influence of the finite size effect and the finite rate effect. The results for the size of 26.6 {\AA} $\times$ 21.9 {\AA} are shown as the red curves in Figure \ref{fig:fig2}(a), and results for other sizes are shown in Figure S2 in the Supporting Information. The results confirm that different pressurization rates do not significantly affect the above conclusion between graphene size and critical crystallization pressure, which means that the variation of critical crystallization pressure for the formation of square ice is not caused by the finite size effect. From the top views of the final simulation snapshots shown in Figure \ref{fig:fig2}(c), the water molecules all form the same NSI (nearly square icing) pattern, even though they are confined between different sizes of the graphene sheets. The results of the hysteresis simulation are shown in Figure S4 in the Supporting Information.

\subsection{Variation of critical crystallization pressure and freezable region for different layers of square ice}
\begin{figure}
  \centering
  \includegraphics[width=0.9\columnwidth]{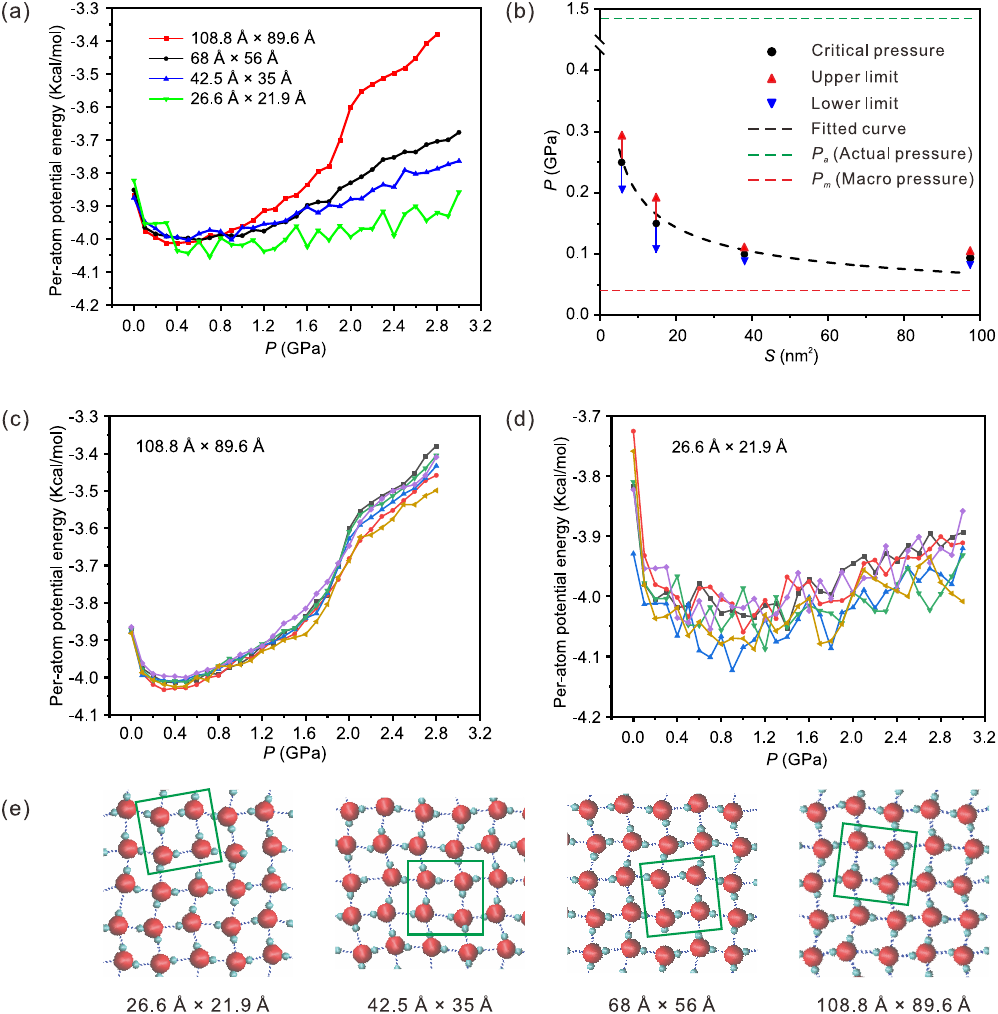}\\
  \caption{Results for the monolayer square ice ($h$ = 6.5 {\AA}). (a) Potential energy curves of the confined water with different graphene sizes. (b) Critical crystallization pressure as a function of graphene size. (c) Variations of the potential energy of the confined water for repeated simulations with different initial velocities when the size of graphene sheets is 108.8 {\AA} $\times$ 89.6 {\AA}. (d) Variations of the potential energy of the confined water for repeated simulations with different initial velocities when the size of graphene sheets is 26.6 {\AA} $\times$ 21.9 {\AA}. (e) Top views of the simulation snapshots for graphene sheets of different sizes.}\label{fig:fig3}
\end{figure}

Although the size of the graphene sheet in $x$ and $z$ direction is varied over a wide range and meaningful conclusions have been summarized for bilayer square ice, the size of nanocapillaries in the capillary width ($y$) direction is still fixed. To further analyze the effect of nanocapillary size on the variation of the critical crystallization pressure, we perform additional MD simulations with different sizes for monolayer ($h$ = 6.5 {\AA}), trilayer ($h$ = 11.5 {\AA}), and four-layer ($h$ = 14.0 {\AA}) ice. When the separation distance between the two graphene sheets is set to be 6.5 {\AA} to accommodate one layer of water molecules, the accurate result of critical crystallization pressure accompanied by a sudden drop of the potential energy is not easy to obtain directly from the diagram of the potential energy, as shown in Figure \ref{fig:fig3}(a). Nevertheless, we can still find an obvious dependence between the critical crystallization pressure and the graphene size, and the critical crystallization pressure increases as the size of the graphene sheet decreases. The lateral pressure $P$ was increased stepwise at a fixed rate of 0.10 GPa every 1 ns in the pressurization process of Figure \ref{fig:fig3}(a), and we also performed simulations with different pressurization rates for comparison, which show similar results, as seen in Figure S5 in the Supporting Information. We repeat the simulations six times for each graphene size by changing the initial velocity distributions of the water molecules while keeping the other settings identical, and count all the results of critical crystallization pressure by observing the ice formation from simulation snapshots. Figure \ref{fig:fig3}(b) summarizes the critical crystallization pressure as a function of the graphene size. As the graphene sheet becomes larger, the critical crystallization pressure slowly decreases, converging to an approximately macroscopic crystallization pressure for monolayer 2D ice ($P_m$ in Figure \ref{fig:fig3}(b)). When we repeat these simulations for each graphene size by changing the initial velocity distributions of the water molecules, the phenomenon of fluctuation for critical crystallization pressure also occurs. The results for two of these graphene sizes are shown in Figure \ref{fig:fig3}(c) and (d). When the size of the graphene sheet is 108.8 {\AA} $\times$ 89.6 {\AA}, even though all curves of potential energy do not coincide after freezing, they drop suddenly at almost the same pressure, showing the consistency of critical crystallization pressure for large graphene sheets. However, when the size of the graphene sheet is minuscule (26.6 {\AA} $\times$ 21.9 {\AA}), the curves of potential energy drop at different pressures, indicating the fluctuation of the critical crystallization pressure. Results for other sizes are shown in Figure S6 in the Supporting Information for details. Figure \ref{fig:fig3}(b) also summarizes the fluctuation of critical crystallization pressure as a function of graphene size. The red and blue arrows represent the upper and lower limits of the critical crystallization pressure, which are the maximum and minimum critical crystallization pressure from six repeated simulations (the distribution of critical crystallization pressure for all cases is shown in Figure S6). These upper and lower limits indicate the fluctuation range of the critical crystallization pressure. The fluctuation range narrows as the sheets widen, converging to the macroscopic behavior of a single value of the critical crystallization pressure for large sheets.

In the same way, the actual pressure of water molecules for monolayer ice ($P_a \approx 1.5$ GPa in Figure \ref{fig:fig3}(b)) can also be roughly estimated from the energy gain due to the squeezing effect. Compared to the water molecules for $h$ = 9.0 {\AA}, the actual pressure $P_a$ is higher and the quasi-macroscopic pressure $P_m$ is lower for $h$ = 6.5 {\AA}. Therefore, the unfreezable threshold for the graphene size is smaller and the freezable region between $P_a$ and $P_m$ is much wider, which means the formation of 2D ice is easier with a narrower width of the capillary and a larger size of the confinement surface. As for the structure of monolayer ice, Figure \ref{fig:fig3}(e) shows the top view snapshots of the results for the width of the capillary $h$ = 6.5 {\AA}, and the water molecules all form the same NSI pattern for different graphene sizes.

\begin{figure}
  \centering
  \includegraphics[width=0.9\columnwidth]{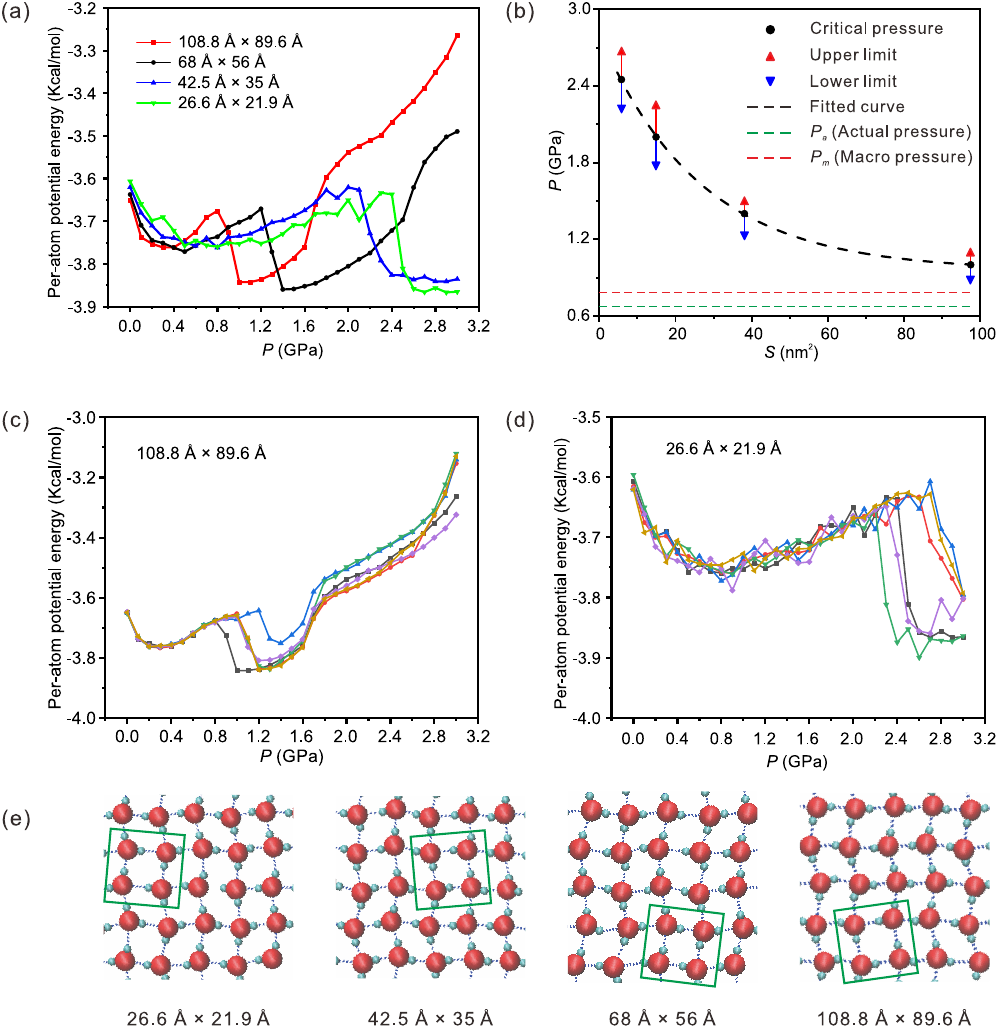}\\
  \caption{Results for trilayer square ice ($h$ = 11.5 {\AA}). (a) Potential energy curves of the confined water with different graphene sizes. (b) Critical crystallization pressure as a function of graphene size. (c) Variations of the potential energy of the confined water for repeated simulations with different initial velocities when the size of graphene sheets is 108.8 {\AA} $\times$ 89.6 {\AA}. (d) Variations of the potential energy of the confined water for repeated simulations with different initial velocities when the size of graphene sheets is 26.6 {\AA} $\times$ 21.9 {\AA}. (e) Top views of the simulation snapshots for graphene sheets of different sizes.}\label{fig:fig4}
\end{figure}

When the separation distance between the two graphene sheets is set to 11.5 {\AA} to accommodate three layers of water molecules, the results of critical crystallization pressure for different graphene sizes are easy to obtain directly from the diagram of the potential energy, and the transition from liquid water to ice is accompanied by a sudden drop of the potential energy curve, as shown in Figure \ref{fig:fig4}(a). As the width of the capillary increases, the critical crystallization pressure increases significantly, and it requires a pressure of nearly 2.5 GPa for the formation of square ice when the size of graphene sheets is minuscule (26.6 {\AA} $\times$ 21.9 {\AA}). There is also obvious dependence between the critical crystallization pressure and the graphene size, and the critical crystallization pressure increases as the size of the graphene sheet decreases. The lateral pressure P was increased with the same rate of 0.10 GPa every 1 ns in the pressurization process of Figure \ref{fig:fig4}(a), and we also performed simulations with different pressurization rates for comparison, which show similar results, as seen in Figure S7 in the Supporting Information for details. We repeat these simulations six times for each graphene size by changing the initial velocity distributions of the water molecules while keeping the other settings identical and counting all the results of critical crystallization pressure. Figure \ref{fig:fig4}(b) summarizes the critical crystallization pressure as a function of the graphene size. As the graphene sheet becomes larger, the critical crystallization pressure slowly decreases, converging to an approximately macroscopic crystallization pressure for trilayer 2D ice ($P_m$ in Figure \ref{fig:fig4}(b)). When we repeat these simulations for each graphene size by changing the initial velocity distributions of the water molecules, the phenomenon of fluctuation for the critical crystallization pressure is even more obvious. The results for two of these graphene sizes are shown in Figure \ref{fig:fig4}(c) and (d). When the size of the graphene sheet is 108.8 {\AA} $\times$ 89.6 {\AA}, all curves of potential energy almost coincide in the liquid phase. However, the curves drop suddenly at different pressures, which means the obvious fluctuation of critical crystallization pressure, even at the large size of the graphene sheet. For the minuscule graphene size (26.6 {\AA} $\times$ 21.9 {\AA}), the critical crystallization pressure fluctuates within a larger range. Results for other sizes are shown in Figure S8 in the Supporting Information for details. Figure \ref{fig:fig4}(b) also summarizes the fluctuation of the critical crystallization pressure as a function of graphene size. The fluctuation range of the critical crystallization pressure is indicated by red and blue arrows, and it narrows as the sheets widen, also converging to the macroscopic behavior of a single value of critical crystallization pressure for large sheets. Comparing the fluctuation range of critical crystallization pressure for trilayer square ice to monolayer square ice, the results are much different from the finite size effect convergence. When the size of graphene sheets is 108.8 {\AA} $\times$ 89.6 {\AA} and 68.0 {\AA} $\times$ 56.0 {\AA}, there is no fluctuation of critical crystallization pressure for monolayer square ice, while the fluctuation is obvious for trilayer square ice, which has a larger size of confinement.

The actual pressure of water molecules for trilayer ice ($P_a \approx 0.6$ GPa in Figure \ref{fig:fig4}(b)) can also be roughly estimated. Compared to the water molecules for $h$ = 6.5 {\AA} and 9.0 {\AA}, the actual pressure $P_a$ is much lower while the macroscopic pressure $P_m$ is higher for $h$ = 11.5 {\AA}. Therefore, the unfreezable threshold for the graphene size is very large and the freezable region between $P_a$ and $P_m$ becomes very narrow, which means the formation of 2D ice is difficult with a large width of the nanocapillary. The results are consistent with that the 2D ice has a strict limit on the number of layers, and can keep stable only for a small range of nanocapillary width\cite{Zangi2004WaterConfinedReview}. As for the structure of trilayer ice, Figure \ref{fig:fig4}(e) shows the top view snapshots of the simulation results for the width of the capillary $h$ = 11.5 {\AA}, and the water molecules all form the same nearly square icing (NSI) pattern for different graphene sizes.

Finite-size effect is an important issue for molecular dynamics calculations of critical properties, and may have contributed to the variation and oscillation of critical crystallization pressure. This effect has been discussed in detail in our previous research\cite{Zeng2023CriticalCrystallizationPressure}. In a finite system, the phase transition may be smeared over a pressure region, and the center of the pressure region may also be shifted, corresponding to the `rounding exponent' and the `shift exponent' \cite{binder1987finite}. In this study, the variations of the critical crystallization pressure correspond to the `shift exponent', and the oscillations of the critical crystallization pressure correspond to the `rounding exponent'. The `rounding exponent' and the `shift exponent' can be presented by many repeated simulation cases. In addition to the three methods used in this study, the critical crystallization pressure can also be obtained by some alternative methods, such as thermodynamic integration along a hypothetical path, which yields the relative free energies of the solid and liquid phases \cite{mastny2007melting}. In a previous study of the solid-liquid melting temperature, it has been shown that finite-size effects can account for deviations in the melting temperature (from the infinite-size limit) of up to 5$\%$ \cite{mastny2007melting}. The deviations in the critical crystallization pressure of square ice (from the infinite-size limit) are more than 20$\%$ in our study, which indicates that the finite-size effects account for only a small part of the pressure variation phenomenon.

\begin{figure}
  \centering
  \includegraphics[width=0.9\columnwidth]{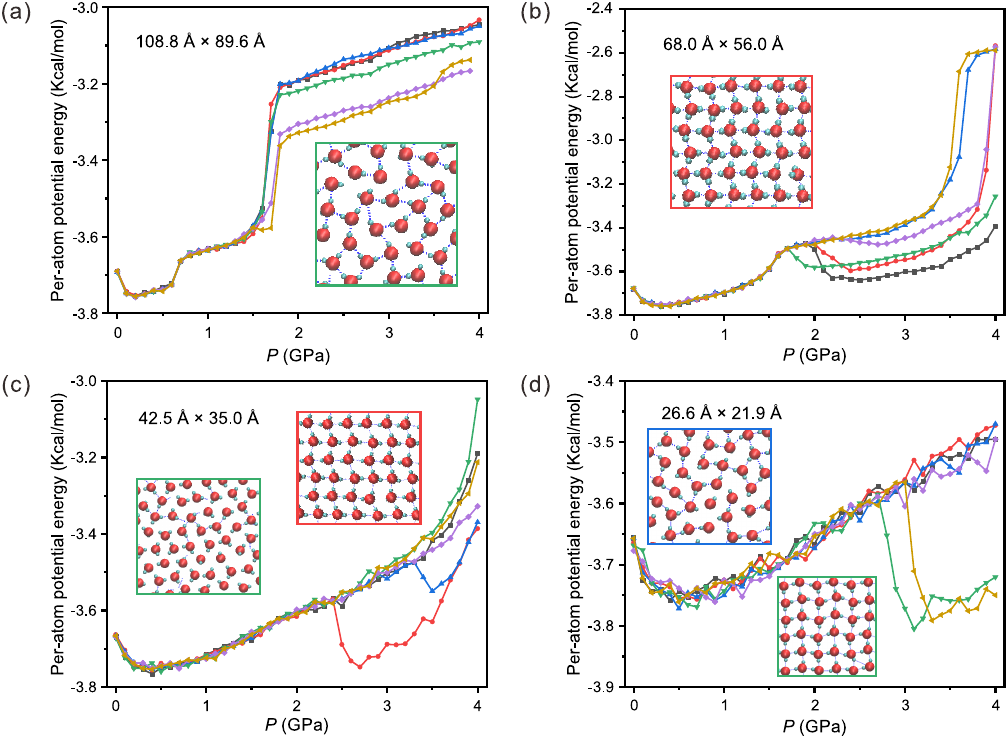}\\
  \caption{Results for four-layer square ice ($h$ = 14.0 {\AA}). (a) 108.8 {\AA} $\times$ 89.6 {\AA}. (b) 68.0 {\AA} $\times$ 56.0 {\AA}. (c) 42.5 {\AA} $\times$ 35.0 {\AA}. (d) 26.6 {\AA} $\times$ 21.9 {\AA}. The edge colors of the inset snapshots correspond to the curves of the potential energy.}\label{fig:fig5}
\end{figure}

Since the freezable region has become very narrow for trilayer square ice as shown in Figure \ref{fig:fig4}(b), the limit of the stable multilayer ice should be close to this region. When we continue to increase the width of the nanocapillary, flat four-layer structures of water molecules are formed as the separation distance between the two graphene sheets is set to be 14.0 {\AA}. However, only some sizes of graphene sheets still have a sudden drop in the potential energy, which corresponds to the critical crystallization pressure for the phase transition, as shown in Figure \ref{fig:fig5}. In contrast, when the size of the graphene sheet is 108.8 {\AA} $\times$ 89.6 {\AA}, the curves of potential energy keep rising during the pressurization process, and no square ice structure is formed, as shown in Figure \ref{fig:fig5}(a). When the size of the graphene sheet is 68.0 {\AA} $\times$ 56.0 {\AA}, even though there is a sudden drop in the potential energy curve, the configuration of water molecules tends to be rhombus rather than square, as shown in Figure \ref{fig:fig5}(b). When the size of the graphene sheet decreases further, we can find the same 2D square ice structures as the monolayer, bilayer, and trilayer square ice, and a similar dependence between the critical crystallization pressure and the graphene size. We repeated these simulations six times for each graphene size by changing the initial velocity distributions of the water molecules while keeping the other settings identical. The results show that even in the same confinement (the same graphene size and the same capillary width), the appearance of four-layer square ice becomes indeterminate. For instance, when the size of the graphene sheet is 26.6 {\AA} $\times$ 21.9 {\AA}, water molecules form a clear square ice structure in two cases, but they are always disordered in the other four cases, as shown in Figure \ref{fig:fig5}(d). Therefore, $h$ = 14.0 {\AA} is the upper limit of the nanocapillary width for the formation of 2D square ice.

\subsection{Limit of the nanocapillary width and correlation to macroscopic crystallization pressure}

\begin{figure}[t]
  \centering
  \includegraphics[width=0.6\columnwidth]{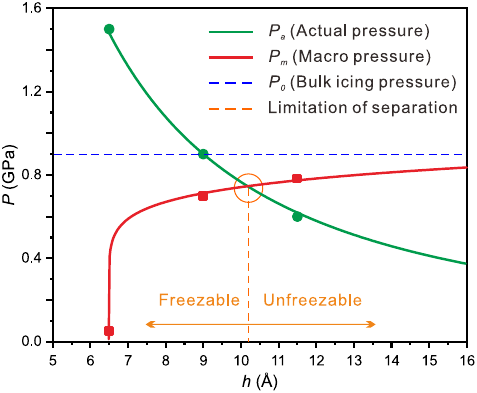}\\
  \caption{Actual pressure of water molecules and quasi-macroscopic crystallization pressure as a function of nanocapillary width. The green and red dots represent the actual pressure of water molecules and the quasi-macroscopic crystallization pressure, respectively. The green and red curves are the fitted curves of the actual pressure and the quasi-macroscopic crystallization pressure, respectively. The blue dashed line represents the crystallization pressure of bulk ice. The orange dashed line represents the limit of the nanocapillary width for the formation of 2D square ice.}\label{fig:fig6}
\end{figure}

Summarizing all simulation results for different sizes of graphene sheets and widths of the capillary, a quasi-macroscopic crystallization pressure and an actual pressure of water molecules can be obtained when we choose a fixed value of nanocapillary width. The quasi-macroscopic crystallization pressure is the minimum critical crystallization pressure for the formation of square ice and is also the minimum pressure required for crystallization with a certain nanocapillary width. The actual pressure of water molecules is the pressure only from the energy gain due to the squeezing effect, without extra external force. The unfreezable threshold means it is hard to crystallize spontaneously in practice when the graphene sheet is smaller than the threshold, because the actual pressure is not high enough to match the critical crystallization pressure of icing. To further unveil the variation rule of critical crystallization pressure for the formation of square ice in graphene nanocapillaries, we summarize all the quasi-macroscopic crystallization pressure and actual pressure of water molecules for different widths of nanocapillaries, as shown in Figure \ref{fig:fig6}. As the width of the capillary becomes larger, the actual pressure decreases markedly, due to the weakening of the squeezing effect. However, the quasi-macroscopic crystallization pressure slowly increases, and the extension of this curve represents the true macroscopic crystallization pressure (approaching macroscale in all three directions), converging to the crystallization pressure of bulk ice\cite{reinhardt2021quantum} ($P_0$ in Figure \ref{fig:fig6}, and the phase diagram for the classical system described by the machine-learning potentials (MLPs), the ab initio phase diagram including nuclear quantum effects (NQEs), as well as the experimental phase diagram). These results establish a direct correlation between 2D square ice and 3D bulk ice with the critical crystallization pressure. The intersection of two pressure curves is the limit of the nanocapillary width for the formation of 2D square ice, which means that it is hard to crystallize spontaneously in practice when the nanocapillary width exceeds the limit, unless additional external pressure is provided. This result can explain why the 2D ice has a strict limit on the number of layers, and can keep stable only for a small range of nanocapillary width\cite{Zangi2004WaterConfinedReview}. It is worth noting that the correlation between 2D square ice and 3D bulk ice by critical crystallization pressure does not mean that we can directly observe the dynamic transformation process between them.

\section{Conclusions}
The variation mechanism of the critical crystallization pressure for 2D square ice formation with the nanocapillary size in all three directions is unveiled, which can explain the limit of the nanocapillary width and establishes a direct correlation between 2D square ice and 3D bulk ice by critical crystallization pressure. The size of the nanocapillary is changed over a wide range in all three directions during the simulation. The results show that there is a strong dependence of the critical crystallization pressure on the size of the confinement sheet, which exists and is similar for monolayer, bilayer, and trilayer ices. The quasi-macroscopic crystallization pressure, actual pressure of water molecules, and freezable region can be obtained for a fixed nanocapillary width, and they are all strongly dependent on the change of the nanocapillary width. As the width of the capillary becomes larger, the actual pressure decreases markedly due to the weakening of the squeezing effect. However, the quasi-macroscopic crystallization pressure slowly increases, converging to the macroscopic crystallization pressure, which is very close to the crystallization pressure for bulk ice. By using critical crystallization pressure, a direct correlation is established between 2D square ice and 3D bulk ice. There is an upper limit of the nanocapillary width for the formation of 2D square ice when the quasi-macroscopic crystallization pressure is equal to the actual pressure. It is hard to crystallize spontaneously in practice when the nanocapillary width exceeds the limit, unless extra external pressure is provided. This can explain why the 2D ice has a strict limit on the number of layers, and can remain stable only for a small range of nanocapillary width. This study provides not only a further understanding of the formation of 2D ice but also physical insights into the connection between 2D and 3D ice.



\section{ASSOCIATED CONTENT}
\subsection{Supporting Information}
The Supporting Information is available free of charge at https://pubs.acs.org/doi/10.1021/acs.jpcc.4c03095
Further details of simulation results with different pressurization rates, variations of potential energy for different sizes of graphene sheets, and different widths of nanocapillary (PDF)

\section{AUTHOR INFORMATION}
\subsection{Corresponding Authors}
Zhizhao Che - State Key Laboratory of Engines \& National Industry-Education Platform of Energy Storage, Tianjin University, Tianjin 300350, China; Email: chezhizhao@tju.edu.cn\\
Tianyou Wang - State Key Laboratory of Engines \& National Industry-Education Platform of Energy Storage, Tianjin University, Tianjin 300350, China; Email: wangtianyou@tju.edu.cn
\subsection{Authors}
Zhen Zeng - State Key Laboratory of Engines, Tianjin University, Tianjin 300350, China\\
Kai Sun - State Key Laboratory of Engines, Tianjin University, Tianjin 300350, China\\
Rui Chen - State Key Laboratory of Engines, Tianjin University, Tianjin 300350, China\\
Mengshan Suo - State Key Laboratory of Engines, Tianjin University, Tianjin 300350, China
\subsection{Author Contributions}
The manuscript was written through contributions of all authors. Zhen Zeng analyzed the data and wrote the manuscript. Kai Sun, Rui Chen, Mengshan Suo, and Zhizhao Che reviewed the manuscript. Tianyou Wang conceptualized the project.

\subsection{Notes}
The authors declare no competing financial interest.

\begin{acknowledgement}
This work was financially supported by the National Natural Science Foundation of China (No.\ 51920105010 and 52176083).
\end{acknowledgement}



\bibliography{Manuscript}


\begin{tocentry}
\centering
\includegraphics{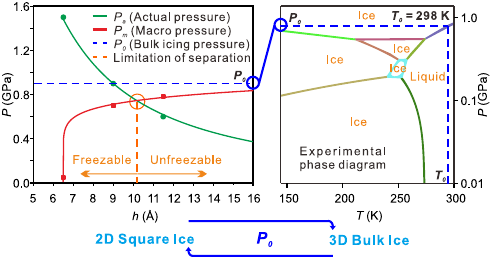}\\
\vspace{3mm}
\end{tocentry}

\end{document}